\documentclass[submission, Proceedings]{SciPost}

\usepackage{amssymb}
\usepackage{amsmath}

\newcommand{\be}{\begin{equation}}
\newcommand{\ee}{\end{equation}}
\newcommand{\ba}{\begin{eqnarray}}
\newcommand{\ea}{\end{eqnarray}}
\newcommand{\la}{\label}
\newcommand{\<}{\langle}
\renewcommand{\>}{\rangle}

\newcommand{\bi}{\begin{itemize}}
\newcommand{\ei}{\end{itemize}}

\newcommand{\amuhvp}{a_\mu^{\rm hvp}}
\newcommand{\amuhlbl}{a_\mu^{\rm hlbl}}

\begin{document}

\begin{center}{\Large \textbf{
Hadronic light-by-light scattering in the anomalous magnetic moment of the muon
}}\end{center}

\begin{center}
N.\ Asmussen\textsuperscript{1,3}, A.\ G\'erardin\textsuperscript{1,4}, A.\ Nyffeler\textsuperscript{1}, H.\ B.\ Meyer\textsuperscript{1,2,*}
\end{center}

\begin{center}
{\bf 1} PRISMA Cluster of Excellence \& Institut f\"ur Kernphysik,
Johannes~Gutenberg-Universit\"at Mainz,  55099 Mainz, Germany
\\
{\bf 2} Helmholtz Institut Mainz, 55099 Mainz, Germany
\\
{\bf 3} School of Physics and Astronomy, University of Southampton, Southampton~SO17~1BJ, United Kingdom
\\
{\bf 4} John von Neumann Institute for Computing, DESY, Platanenallee~6, D-15738~Zeuthen, Germany
\\
* Speaker; email: meyerh@uni-mainz.de
\end{center}

\begin{center}
\today
\end{center}

\definecolor{palegray}{gray}{0.95}
\begin{center}
\colorbox{palegray}{
  \begin{tabular}{rr}
  \begin{minipage}{0.05\textwidth}
    \includegraphics[width=8mm]{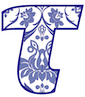}
  \end{minipage}
  &
  \begin{minipage}{0.82\textwidth}
    \begin{center}
    {\it Proceedings for the 15th International Workshop on Tau Lepton Physics,}\\
    {\it Amsterdam, The Netherlands, 24-28 September 2018} \\
    \href{https://scipost.org/SciPostPhysProc.1}{\small \sf scipost.org/SciPostPhysProc.Tau2018}\\
    \end{center}
  \end{minipage}
\end{tabular}
}
\end{center}


\section*{Abstract}
{\bf
Hadronic light-by-light scattering in the anomalous magnetic moment of the muon $a_\mu$
is one of two hadronic effects limiting the precision of the Standard Model prediction
for this precision observable, and hence the new-physics discovery potential of 
direct experimental determinations of $a_\mu$.
In this contribution, we report on recent progress in the calculation of this effect achieved both via dispersive 
and lattice QCD methods.
}

\vspace{10pt}
\noindent\rule{\textwidth}{1pt}
\noindent\rule{\textwidth}{1pt}
\vspace{10pt}

\section{Introduction}
\label{sec:intro}

The magnetic moment of the muon is one of the most precisely measured
quantities in particle physics.  In units of
$\frac{e}{2m_\mu}\cdot\frac{\hbar}{2}$, its value is given by the
gyromagnetic factor $g$.  The prediction that $g=2$ was an early
success of the Dirac equation, applied to the electron.  
The relative deviation of the gyromagnetic factor from the Dirac prediction
is conventionally called anomalous magnetic moment, and is denoted by $a_\mu\equiv (g-2)_\mu/2$.
Remarkably, the quantity $a_\mu$ has been directly measured to 0.54ppm of
precision~\cite{Bennett:2006fi}.  The Standard Model (SM)
prediction for $a_\mu$, see e.g.\ \cite{Jegerlehner:2009ry}, is currently at a
similar precision level, 0.37ppm~\cite{Davier:2017zfy}.
The precision of the SM prediction is entirely limited by the hadronic contributions.
Specifically, the hadronic vacuum polarisation, which enters at O($\alpha^2$), 
and the hadronic light-by-light contribution $\amuhlbl$, which is of order $\alpha^3$, 
contribute in comparable amounts to the absolute uncertainty;
their respective depiction as Feynman diagrams is shown in Fig.\ \ref{fig:Feyndiag}.
The new E989 experiment at Fermilab is underway (see~\cite{Fertl:2016nij} 
and the presentation of A.\ Driutti at this conference), with the stated goal of improving
the precision of the measurement by a factor four,  and the E34 experiment at J-PARC 
(see~\cite{Otani:2015lra} and the presentation of T.\ Mibe at this conference)
plans to achieve a similar precision with a very different technique.
It is therefore essential to improve the precision of the predictions for the hadronic contributions
in order to enhance the new-physics sensitivity of the upcoming experimental results. 

In view of the observations above, the theory precision requirements for the short-term future are the following: 
for the hadronic vacuum polarisation contribution, $\amuhvp\approx 6900\cdot10^{-11}$, 
the goal is to consolidate the currently quoted~\cite{Keshavarzi:2018mgv} precision of 0.35\% obtained using 
a dispersive representation with experimental $e^+e^-$ data as input, 
and to approach that level of precision in lattice calculations~\cite{Meyer:2018til};
while for $\amuhlbl\approx 100\cdot 10^{-11}$ a precision of 10 to 15\% suffices.
Clearly, both tasks are very challenging. In this talk, we focus on $\amuhlbl$ and refer the reader
to the talk of Ch.\ Lehner for the status of $\amuhvp$.

The activities linked to the determination of $\amuhlbl$ can be divided into four classes:
\begin{enumerate}
\item {\bf Model calculations}, which constituted the only approach until 2014, are based on pole- and loop-contributions of hadron resonances,
in some cases also on constituent quark loops.
\item {\bf Dispersive approaches} allow one to identify and compute individual hadronic contributions in terms of physical observables, such 
as transition form factors and $\gamma^*\gamma^*\to\pi\pi$ amplitudes.
\item A dedicated {\bf experimental program} is needed to provide input for the model \& dispersive approaches, e.g.\ 
$(\pi^0,\eta,\eta')\to \gamma\gamma^*$ at virtualities $Q^2\lesssim 3\,${GeV}$^2$; there is in particular an active program at BES-III on this theme, 
see the talk by Y.\ Guo at this conference.
\item In terms of {\bf lattice calculations}, two groups (RBC/UKQCD and Mainz) have been working on formulating and carrying out a direct 
lattice calculation of $\amuhlbl$.
\end{enumerate}
An important question is ultimately, how well the findings from the different approaches fit together.
We begin by reviewing aspects of the model calculations and describing how one can test the assumptions underlying them;
carry on to describe the status of the dispersive approaches and finally discuss in more detail several aspects 
of the lattice calculations of $\amuhlbl$.

\begin{figure}[t]
\centerline{\includegraphics[width=0.36\textwidth]{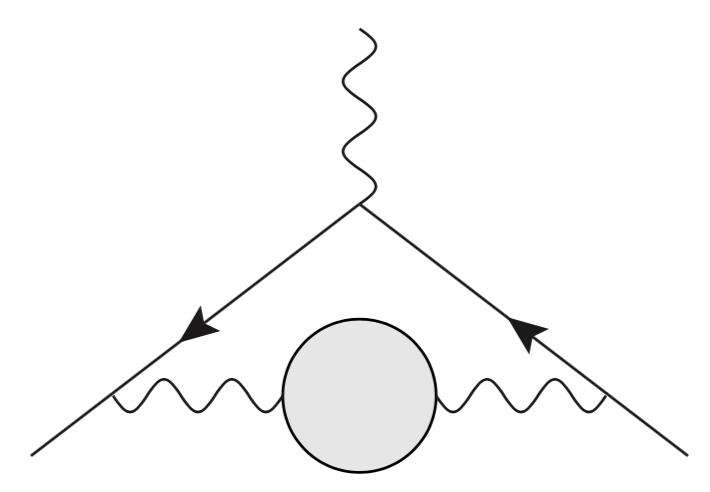}
~~~~~~~~~~~~\includegraphics[width=0.26\textwidth]{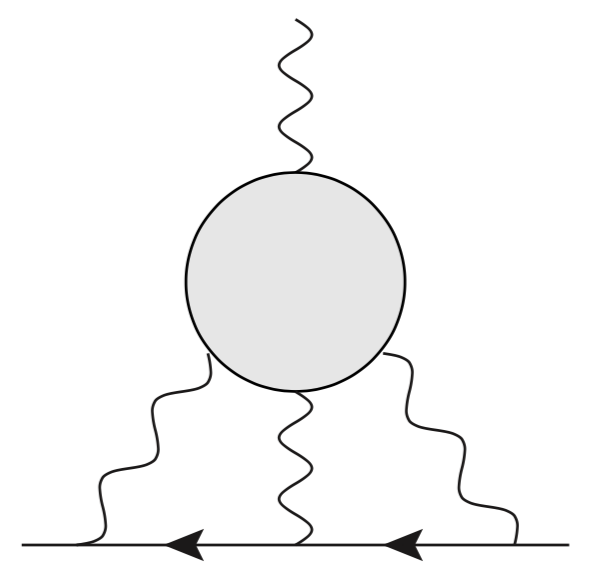}}   
\caption{\la{fig:Feyndiag}The hadronic contributions to $(g-2)_\mu$ dominating the theory uncertainty budget.
Left: the hadronic vacuum polarisation contribution. Right: the hadronic light-by-light scattering contribution.
A solid line represents the muon propagator, the wavy lines represent photon propagators. The external magnetic field 
is represented by a photon line coming in from the top.}
\end{figure}

\section{Learning from, and testing hadronic models \la{sec:modeltest}}
A recently updated estimate from the hadronic model calculations is~\cite{Jegerlehner:2018gjd}
\be
\amuhlbl = (103 \pm 29) \cdot 10^{-11}.
\ee
As compared to earlier estimates, the pole contribution of the axial-vector mesons has been revised and is much smaller.
Nevertheless, the central value of the estimate has changed little since the 2009 `Glasgow consensus'
estimate of $(105\pm26)\cdot 10^{-11}$~\cite{Prades:2009tw}.
Beyond the numerical result of the model calculations, it is worth recording some of the physics lessons
learnt from them~\cite{Prades:2009tw}:
\bi
\item A heavy (charm) quark loop makes a small contribution
\[
\amuhlbl = \left(\frac{\alpha}{\pi}\right)^3\, N_c {\cal Q}_c^4 c_4  \frac{m_\mu^2}{m_c^2}, \qquad c_4\approx 0.62,
\]
where ${\cal Q}_c$ and $m_c$ are respectively the charm quark charge and mass, $N_c=3$ is the number of colors, $m_\mu$ is the muon mass
and $\alpha\approx 1/137$ is the fine-structure constant.
\item For light-quarks, the most relevant degrees of freedom are the pions. 
The leading contribution in chiral perturbation theory, namely the charged-pion loop calculated in scalar QED, depends only on $m_\mu/m_\pi$, with 
\\ 
\be\la{eq:piloop}
\amuhlbl \stackrel{m_\mu\ll m_\pi}{=\!=\!=} \left(\frac{\alpha}{\pi}\right)^3\, c_2\frac{m_\mu^2}{m_\pi^2},\qquad c_2\approx -0.065.
\ee
Numerically, this contribution amounts to $\amuhlbl \approx -45\cdot10^{-11}$ for the physical value of $m_\mu/m_\pi$.
Secondly, the neutral-pion exchange is positive and sensitive to the confinement scale~\cite{Knecht:2001qf,Knecht:2001qg},
\be \la{eq:pi0}
\amuhlbl = \left(\frac{\alpha}{\pi}\right)^3\,  \frac{N_c^2\,m_\mu^2}{48\pi^2 F_\pi^2}\,\Big[\log^2\frac{m_\rho}{m_\pi} + {\rm O}\Big(\log \frac{m_\rho}{m_\pi}\Big)
+ {\rm O}(1) \Big].
\ee
We note that, the pion decay constant $F_\pi\approx92\,$MeV being of order $N_c^{1/2}$, 
the contribution (\ref{eq:pi0}) is enhanced by a factor $N_c$ relative to the pion loop, Eq.\ (\ref{eq:piloop}).
On the other hand, the latter is dominant in the limit $m_\pi\to0$ with $m_\mu/m_\pi$ fixed.
The $\rho$ meson mass appears as the hadronic scale regulating an ultra-violet divergence, which appears if one
assumes a virtuality-independent $\pi^0\to\gamma^*\gamma^*$ coupling.

\item For real-world quark masses, using form factors for the mesons is essential in obtaining quantitative results, 
and resonances up to 1.5\,GeV can still be relevant. This makes $\amuhlbl$ sensitive to QCD at intermediate energies,
which is difficult to handle by analytic methods.
\item Some information can be obtained from the operator-product expansion. 
Two closeby vector currents 
\be
V_\mu(x)V_\nu(0)\stackrel{\rm OPE}{\sim} \epsilon_{\mu\nu\rho\sigma}\frac{x_\rho}{(x^2)^2}A_\sigma(0)+\dots
\ee
`look like' an axial current from a distance. For that reason, the doubly-virtual transition form factors of $0^{-+}$ and $1^{++}$ mesons only fall off like $1/Q^2$.
This singles out the poles associated with pseudoscalar and axial-vector mesons as being particularly relevant.
However, the coupling of an axial-vector meson to two \emph{real} photons is forbidden by the Yang-Landau theorem~\cite{Landau:1948kw,Yang:1950rg},
suggesting that the pseudoscalar mesons $\pi^0,\eta,\eta'$ are the most important pole contributions in $\amuhlbl$ due to their 
unsuppressed coupling to two photons, in addition to their relatively light masses.
\ei

The applicability of the hadronic model to $\amuhlbl$ can be tested by 
predicting the relevant four-point correlation function of the electromagnetic current 
$j^\mu = \sum_{f=u,d,s,\dots} {\cal Q}_f \bar q\gamma_\mu q$ and confronting the prediction with non-perturbative lattice QCD data.
The Euclidean momentum-space four-point function at spacelike virtualities can indeed be computed in lattice QCD~\cite{Green:2015sra,Gerardin:2017ryf},
\ba\la{eq:masterl}
&& \Pi_{\mu_1\mu_2\mu_3\mu_4}(P_4;P_1,P_2) \equiv 
\int_{X_1,X_2,X_4} e^{-i \sum_{a} P_a\cdot X_a} \Big\<J_{\mu_1}(X_1)J_{\mu_2}(X_2)J_{\mu_3}(0)J_{\mu_4}(X_4)\Big\>
\ea
and projected to one of the eight forward $\gamma^*\gamma^*\to\gamma^*\gamma^*$ scattering amplitudes, for instance
\ba
&& {\cal M}_{\rm TT}(-Q_1^{\,2},-Q_2^{\,2},-Q_1\cdot Q_2)
= \frac{ e^4}{4} {R_{\mu_1\mu_3}R_{\mu_2\mu_4}}  \Pi_{\mu_1\mu_3\mu_4\mu_2}(-Q_2;-Q_1,Q_1).
\ea
In this particular case, the projectors $R_{\mu\nu}$ project onto the plane orthogonal to the vectors $Q_1$ and $Q_2$
and ${\cal M}_{\rm TT}$ thus corresponds to the amplitude involving transversely polarized photons.
Dispersive sum rules have been derived for the forward amplitudes~\cite{Pascalutsa:2010sj,Pascalutsa:2012pr}. With 
 $\nu = \frac{1}{2}(s+Q_1^2+Q_2^2)$, a crossing-symmetric variable parametrizing the center-of-mass energy $\sqrt{s}$, 
we can write a subtracted dispersion relation,
\be
 {\cal M}_{\rm TT}(q_1^2,q_2^2,\nu) - {\cal M}_{\rm TT}(q_1^2,q_2^2,0) 
  = \frac{2\nu^2}{\pi}
\int_{\nu_0}^\infty d\nu'\frac{ \sqrt{ \nu'{}^2 - q_1^2 q_2^2  }}{\nu'(\nu'{}^2-\nu^2-i\epsilon)} (\sigma_0+\sigma_2)(\nu'),
\ee
where $\sigma_{J}$ corresponds to the total cross-section for the photon-photon fusion reaction $\gamma^*\gamma^*\to{\rm hadrons}$ 
with total helicity $J$.

\begin{figure}[t]
\centerline{\includegraphics[width=0.520\textwidth]{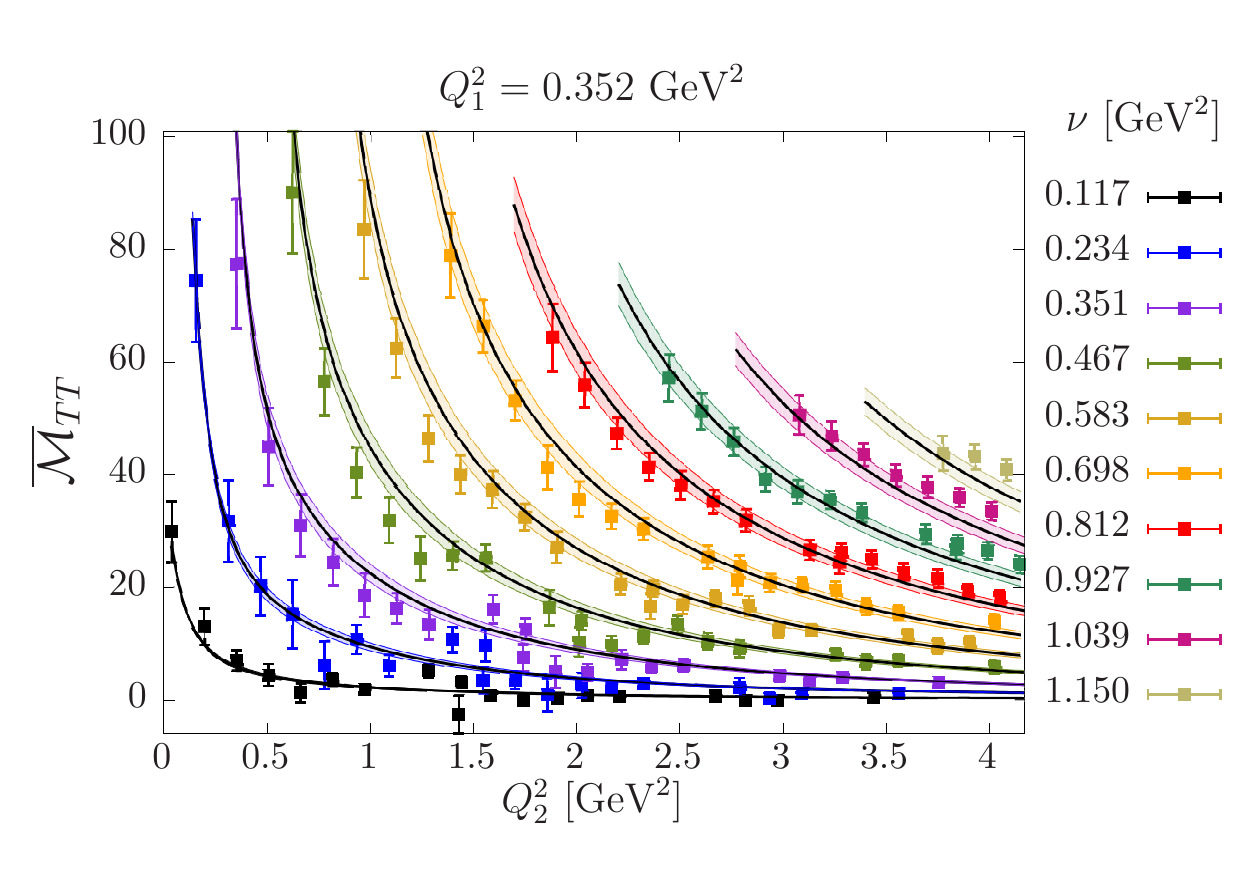}
\includegraphics[width=0.520\textwidth]{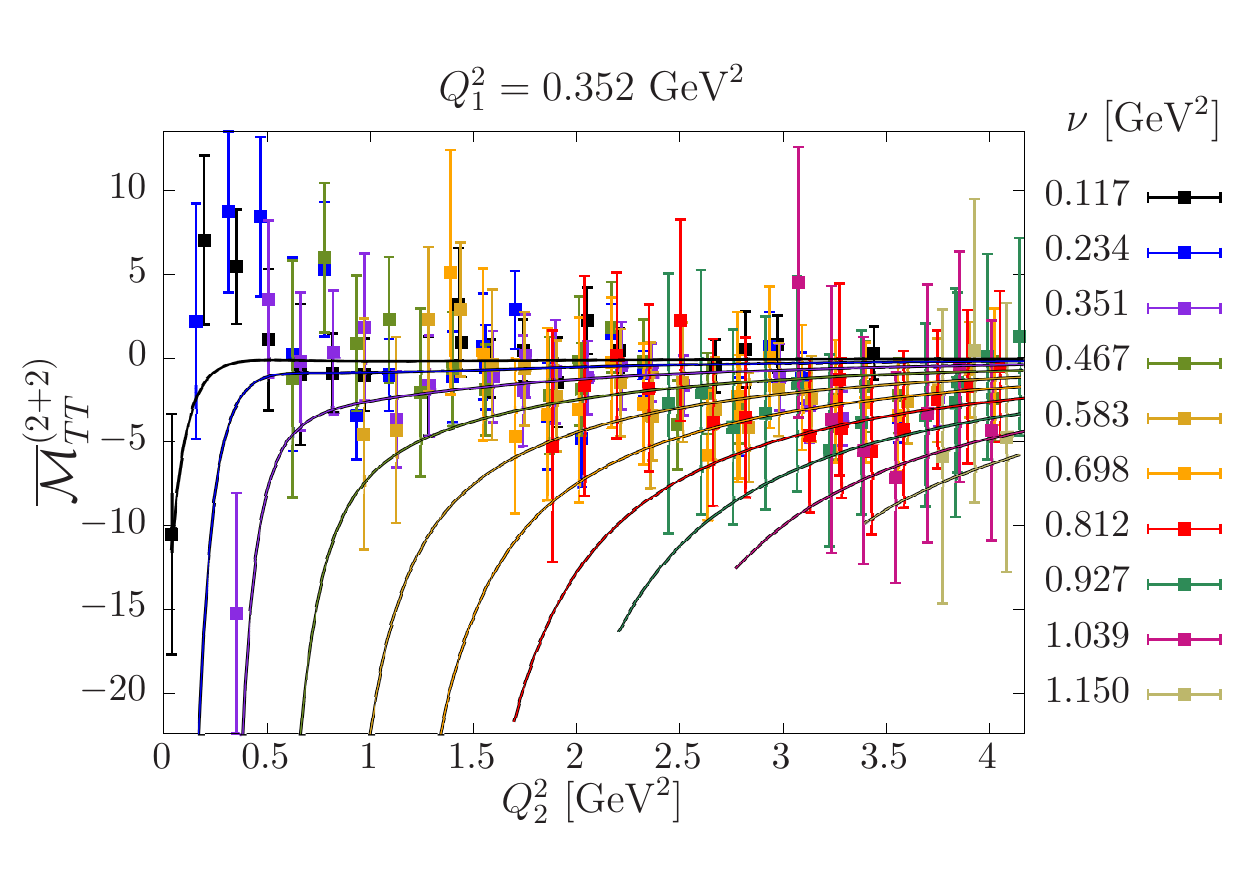}}
\vspace{-0.3cm}
\caption{\la{fig:MTT} The subtracted forward hadronic light-by-light amplitude ${\cal M}_{\rm TT}(-Q_1^2,-Q_2^2,\nu)-{\cal M}_{\rm TT}(-Q_1^2,-Q_2^2,0)$,
multiplied by $10^6$, computed on a $48^3\times96$ lattice ensemble with $m_\pi=314\,$MeV and lattice spacing $a=0.065\,$fm. Left: contribution of the fully connected 
class of quark contractions. Right: contribution of the (2+2) quark-contraction class~\cite{Gerardin:2017ryf}.}
\end{figure}

While experimental data exists for the fusion of real photons into hadrons, no such data is available for spacelike photons. 
In order to model the corresponding cross-section, we note that the contribution of a narrow meson resonance is 
\be
\sigma_{\gamma^*\gamma^*\to{\rm resonance}} 
\propto ~\delta(s-M^2)\times \Gamma_{\gamma\gamma}\times \Big[\frac{{\cal F}_{M\gamma^*\gamma^*}(Q_1^2,Q_2^2)}{{\cal F}_{M\gamma^*\gamma^*}(0,0)} \Big]^2.
\ee
It is then interesting to test whether all eight forward LbL
amplitudes obtained from lattice computations can be described by such
a sum of resonances via the dispersive sum rule. Essential ingredients
in this parametrization of $\sigma_{\gamma^*\gamma^*\to{\rm hadrons}}$
are the transition form factors ${\cal F}_{M\gamma^*\gamma^*}(Q_1^2,Q_2^2)$,
describing the coupling of the resonance to two virtual photons.  In
the case of the neutral pion, a dedicated lattice QCD calculation of
${\cal F}_{\pi^0\gamma^*\gamma^*}$ was performed~\cite{Gerardin:2016cqj},
thus allowing for a definite prediction for this contribution. For the other
included hadronic resonances, which have quantum numbers $J^{PC}=0^{\pm+},~1^{++},~2^{++}$, 
a monopole or dipole parametrization of the virtuality-dependence of the transition form factors was chosen and
fitted to the lattice data for the forward LbL amplitudes. In addition to the resonances, the Born expression for 
$\sigma_{\gamma^*\gamma^*\to\pi\pi}$ was included in the cross-section.
A satisfactory description of the data was obtained in this way; see Fig.\ \ref{fig:MTT}.

The calculation of the four-point correlation function of a quark
bilinear in lattice QCD requires computing the Wick contractions
of the quark fields, since the action is bilinear in these fields.
The major difference with perturbation theory is that the quark
propagators have to be computed in a non-perturbative gauge field
background, which means inverting the sparse matrix of typical size
$10^8\times 10^8$ representing the discretized Dirac operator, on a
source vector. The back-reaction of the quarks on the gauge field is
taken into account in the importance sampling of the gauge fields.
Five classes of Wick contractions contribute to the full four-point
correlation function, as illustrated in Fig.\ \ref{fig:contra}.  While
the fully connected class of diagrams can be computed cost-effectively
using `sequential' propagators, the other classes require the use of
stochastic methods.  In \cite{Gerardin:2017ryf}, only the first two
classes, denoted by the symbols (4) and (2+2), were computed, because
the other three classes (3+1), (2+1+1) and (1+1+1+1) are expected to
yield significantly smaller contributions. If this expectation is
correct, and if the LbL amplitude is dominated by resonance
exchanges, one can infer with what weight factors the isovector and the isoscalar
resonances contribute to the leading contraction topologies (4) and (2+2).
The isoscalar resonances contribute (with unit weight) to the class (2+2);  the isovector resonances
overcontribute with a weight factor $34/9$ to class (4), while the (2+2) contractions compensate
with a weight factor of $-25/9$~\cite{Bijnens:2016hgx}. These counting rules have been used in describing the 
lattice data in Fig.\ \ref{fig:MTT}. In particular, the large-$N_c$ inspired counting rules 
suggest that there is a large cancellation between the isovector resonances and the isoscalar resonances in 
the (2+2) class of diagrams, with the exception of the pseudoscalar mesons, due to the large 
mass difference between the $\pi^0$  and the $\eta'$ meson. Therefore, the contribution of the (2+2) diagrams
to the light-by-light amplitudes was modelled as the $\eta'$ contribution, minus $\frac{25}{9}$ times the $\pi^0$ contribution.
Within the $\sim30\%$ uncertainties, the lattice data was successfully 
reproduced\footnote{In \cite{Gerardin:2017ryf}, it was also shown that in the SU(3) symmetric theory,
similar arguments apply to the contribution of the flavor-octet and singlet mesons, the octet contributing with a weight factor 3 
to the diagram-class (4) and with weight factor of $(-2)$ to diagram-class (2+2), while the singlet only contributes (with unit weight) to diagram-class (2+2).}.

\begin{figure}[t]
\centerline{\includegraphics[width=0.86\textwidth]{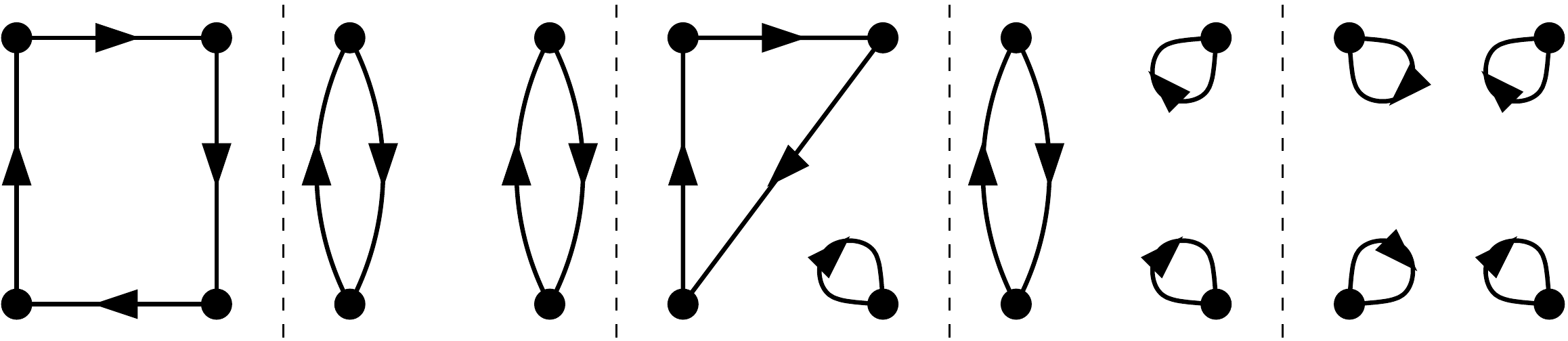}}
\caption{\la{fig:contra} The five classes of quark Wick contractions contributing to the four-point function of the vector current. 
From left to right: class (4), (2+2), (3+1), (2+1+1) and (1+1+1+1). Figure by J.\ Green.}
\end{figure}

Thus the exploratory study \cite{Gerardin:2017ryf} found that the LbL
tensor (\ref{eq:masterl}) at moderate spacelike virtualities can be
described by a set of resonance poles, much in the same way that
$\amuhlbl$ is obtained in the model calculations.  It would be worth
exploring this avenue further, in particular by increasing the
precision of the lattice calculation.

\section{Dispersive approach to $\amuhlbl$ and its input\la{sec:disp}}

Several dispersive approaches have been proposed to handle the complicated physics of 
hadronic light-by-light scattering~\cite{Pauk:2014rfa,Colangelo:2014dfa,Hagelstein:2017obr}.
Here we will mainly review aspects of the `Bern approach'~\cite{Colangelo:2014dfa},
which is the furthest developed at this point in time.
It was shown that the full hadronic light-by-light tensor can be decomposed into 54 Lorentz structures~\cite{Colangelo:2015ama},
\be\la{eq:Pi1}
\Pi^{\mu\nu\lambda\sigma}(q_1,q_2,q_3) = i^3\int_{x,y,z} \!\!\!\!\!\!\! e^{-i(q_1x+q_2y+q_3z)}\<0|T\{j^\mu_x j^\nu_y j^\lambda_z j^\sigma_0\}|0\> 
= \sum_{i=1}^{54} T_i^{\mu\nu\lambda\sigma} \Pi_i.
\ee
The Lorentz-invariant coefficients $\Pi_i$ are entirely determined by seven functions of the invariants $q_i\cdot q_j$ combined with crossing symmetry.
The 54 Lorentz structures are redundant, but they allow one to avoid kinematic singularities.

The HLbL contribution to $(g-2)_\mu$ is then computed using the  projection technique, i.e.\ directly at $q=0$:
\be
\amuhlbl = -e^6 \int \frac{d^4q_1}{(2\pi)^4}\frac{d^4q_2}{(2\pi)^4}
\frac{\sum_{i=1}^{12} \hat T_i(q_1,q_2;p)\,\hat\Pi_i(q_1,q_2,-q_1-q_2)}{q_1^2\, q_2^2\,(q_1+q_2)^2\, [(p+q_1)^2-m_\mu^2]\,[(p-q_2)^2-m_\mu^2]}
\ee
The $\hat\Pi_i$ are linear combinations of the $\Pi_i$ appearing in Eq.\ (\ref{eq:Pi1}). 

Performing all ``kinematic'' integrals using the Gegenbauer-polynomial technique after performing a Wick rotation, 
the expression can be reduced to a three-dimensional integral, 
\be\la{eq:masterP}
\amuhlbl = \frac{2\alpha^3}{3\pi^2}\int_0^\infty \!\!\!\! d|Q_1|\,|Q_1|^3 \!\!\int_0^\infty\!\!\!\! d|Q_2|\,|Q_2|^3\!\!\int_{-1}^1 \!\!d\tau\sqrt{1-\tau^2}
\sum_{i=1}^{12} T_i(|Q_1|,|Q_2|,\tau)\,\bar\Pi_i(|Q_1|,|Q_2|,\tau),
\ee
the master relation in this approach ($\tau= Q_1\cdot Q_2/(|Q_1|\,|Q_2|)$).

The contribution of the pole contributions associated with pseudoscalar mesons was worked out explicitly
and clarified the way that the corresponding transition form factors are to be applied in this framework; see subsection \ref{sec:piTFF} below.
As a further result obtained as part of this approach, it was shown~\cite{Colangelo:2017qdm} that certain contributions
in the dispersive approach to the pion loop could be handled rather accurately,
\be
 a_\mu^{\pi\,{\rm box}} + a_{\mu,J=0}^{\pi\pi,\pi{\rm-pole LHC}}=  -24(1)\cdot 10^{-11}.
\ee
The rescattering effects of the pions are being worked out for partial waves $\ell\leq 2$~\cite{Colangelo:2017fiz};
first results by the Bern group for the $s$-wave were presented at the $(g-2)$ Theory Initiative Workshop~\cite{Colangelo2018w}.
An independent analysis of the $\gamma\gamma^*\to \pi\pi$ process has also appeared very recently~\cite{Danilkin:2018qfn}.

\subsection{The transition form factor of the pion\la{sec:piTFF}}

The field-theoretic definition of the transition form factor of the pion involves 
a time-ordered product of two vector currents,
\be
M_{\mu\nu}(p,q_1)  \equiv i \int \mathrm{d}^4 x \, e^{i q_1 x} \, \langle
\Omega | T \{ j_{\mu}(x) j_{\nu}(0) \} | \pi^0(p) \rangle =
\epsilon_{\mu\nu\alpha\beta} \, q_1^{\alpha} \, q_2^{\beta} \,
{\cal F}_{\pi\gamma^*\gamma^*}(q_1^2, q_2^2),
\ee
with $p=q_1+q_2$.
A detailed dispersive analysis of the $\pi^0\to\gamma^*\gamma^*$ transition form factor has recently been carried out~\cite{Hoferichter:2018dmo}, 
leading to the rather accurate result
\be
 a_\mu^{{\rm HLbL},\pi^0} = 62.6^{+3.0}_{-2.5}\cdot 10^{-11}.
\ee

In addition to experiments, important input for the dispersive approaches can be provided by lattice QCD.
A first calculation of ${\cal F}_{\pi\gamma^*\gamma^*}(q_1^2, q_2^2)$
was carried out in lattice QCD with the two lightest quark
flavors~\cite{Gerardin:2016cqj} and used to calculate $a_\mu^{{\rm  HLbL},\pi^0}$, 
obtaining the result $(65.0\pm8.3) \cdot 10^{-11}$. A model parametrization of the 
transition form factor was used here which incorporates known constraints at 
asymptotically large virtualities\footnote{No use was made of the experimentally accurately known 
normalization, ${\cal F}_{\pi\gamma^*\gamma^*}(0,0)$, from the $\pi^0$ width.}.
A second calculation in QCD, including also the dynamical effects of the strange quark,
and using a more model-independent conformal-mapping parametrization of ${\cal F}_{\pi\gamma^*\gamma^*}(q_1^2,q_2^2)$,
obtained the preliminary result $\amuhlbl|_{\pi^0} = (60.4\pm3.6)\cdot10^{-11}$~\cite{Gerardin2018w}.
The lattice and dispersive results are thus in excellent agreement, and comparable in precision.
It is somewhat surprising how close the central values are to older estimates based on the simplest
vector-meson dominance model of the form factor, 
e.g. $\amuhlbl|_{\pi^0,{\rm VMD}} = 57.0\cdot10^{-11}$ \cite{Knecht:2001qf}; however,
the uncertainty of the result is now much better known.

\section{The direct lattice calculation of HLbL in $(g-2)_\mu$}

The idea to directly calculate $\amuhlbl$ in lattice QCD was pioneered in~\cite{Hayakawa:2005eq}.
At first, the task was thought of as a combined QED\,+\,QCD calculation. Today's viewpoint 
is that the calculation amounts to a QCD four-point function, to be integrated over
with a weighting kernel which represents all the QED parts, i.e.\ muon and photon propagators.
Two collaborations have so far embarked on this challenging endeavour~\cite{Meyer:2018til}.

The RBC/UKQCD collaboration has performed calculations of $\amuhlbl$
using a coordinate-space method in the muon rest-frame. The photon and
muon propagators are either computed on the same $L\times L\times L\times T$
torus as the QCD fields -- this approach goes under the name of
QED$_L$ and was first published in~\cite{Blum:2015gfa}; or they are computed 
in infinite volume in a method called QED$_\infty$~\cite{Blum:2017cer}.

The Mainz group has used a manifestly Lorentz-covariant QED$_{\infty}$
coordinate-space strategy, presented in~\cite{Green:2015mva},
averaging over the muon momentum using the Gegenbauer polynomial
technique. This technique relies on the anomalous
magnetic moment of the muon being a Lorentz scalar quantity; a fact that has been
used extensively in the phenomenology community.

\begin{figure}[t]
\centerline{\begin{minipage}{5.8cm}\vspace{-3.55cm}\centerline{QED kernel $\bar{\cal L}_{[\rho,\sigma];\mu\nu\lambda}(x,y)$}
\smallskip
\includegraphics[width=\textwidth]{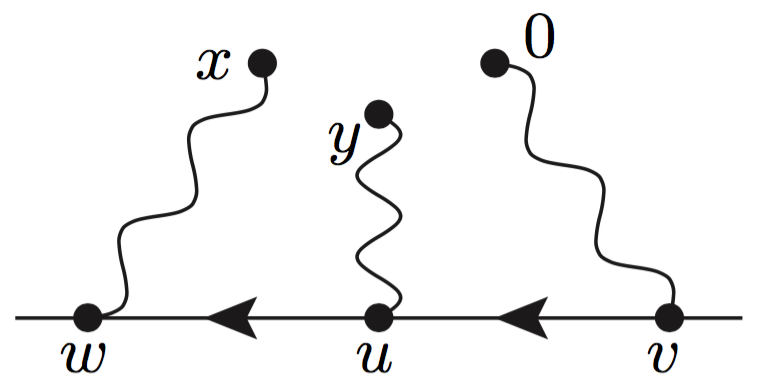}\end{minipage}\begin{minipage}{1.4cm}\vspace{-2cm}
\centerline{\Large$\Rightarrow$}\end{minipage}
\includegraphics[width=0.32\textwidth]{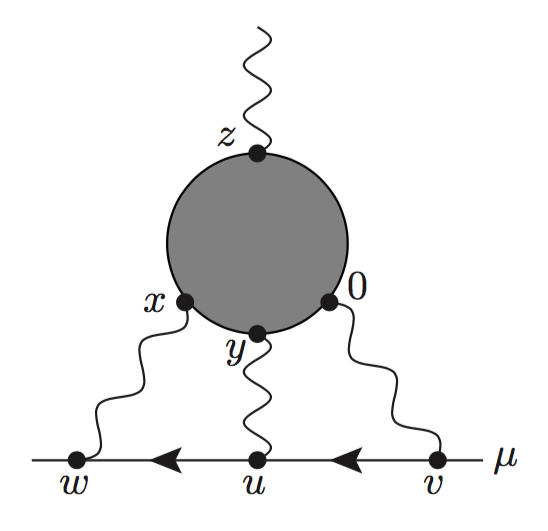}}
\caption{\la{fig:HLbLkernel} Coordinate-space Feynman diagram for $\amuhlbl$, where the QED part, on the left, 
can be computed (semi-)analytically and treated as a kernel weighting the vertex positions of the QCD four-point function of the vector current.}
\end{figure}

A theoretical advantage of using the QED$_{\infty}$ formulation is that 
no power-law finite-volume effects appear, which arise in QED$_L$ due to the massless photon 
propagators\footnote{Instead, the leading finite-size effects are expected to be of order $\exp(-m_\pi L/2)$.}.
Specifically, in the (Euclidean) notation used by the Mainz group, the master equation for computing $\amuhlbl$ is 
\ba
\amuhlbl &=& \frac{m e^6}{3}  \underbrace{\int d^4y}_{=2\pi^2 |y|^3 d|y|}
 \Big[\int d^4x  \underbrace{\bar{\cal L}_{[\rho,\sigma];\mu\nu\lambda}(x,y)}_{\rm QED}\;  
  \underbrace{i\widehat\Pi_{\rho;\mu\nu\lambda\sigma}(x,y)}_{=\;\textrm{ QCD~``blob''}}\Big],
\la{eq:master}\\
i\widehat \Pi_{\rho;\mu\nu\lambda\sigma}( x, y)  &=& 
-\int d^4z\; z_\rho\, \Big\<\,j_\mu(x)\,j_\nu(y)\,j_\sigma(z)\, j_\lambda(0)\Big\>.
\ea
The QED kernel $\bar{\cal L}_{[\rho,\sigma];\mu\nu\lambda}(x,y)$ is
computed in the continuum and in infinite-volume; it consists of the
muon and photon propagators depicted in Fig.\ \ref{fig:HLbLkernel}.
Once the two tensors appearing in Eq.\ (\ref{eq:master}) are
contracted, the integrand is a Lorentz scalar, and the integrals over
$x$ and $y$ reduce to an integral over three invariant variables,
e.g.\ $(x^2,x\cdot y,y^2)$.  In this sense, Eq.\ (\ref{eq:master}) is
the coordinate-space analogue of Eq.\ (\ref{eq:masterP}).  In
practical lattice calculations, one may carry out the four-dimensional
summation over one of the vertices (say $x$) in full, because this
tends to have a beneficial averaging effect. For fixed vertex position
$y$, the four-dimensional summation over $x$ can be arranged to be
performed exactly on every gluon field configuration with an affordable number of 
operations for the most important Wick contraction classes (4) and
(2+2).  Sampling all values of the vertex position $y$ would be
computationally too costly; reducing instead the integral over $y$ to a
one-dimensional integral~\cite{Green:2015mva} allows one to sample the
integrand reliably.

\begin{figure}
\centerline{ \includegraphics[width=0.51\textwidth]{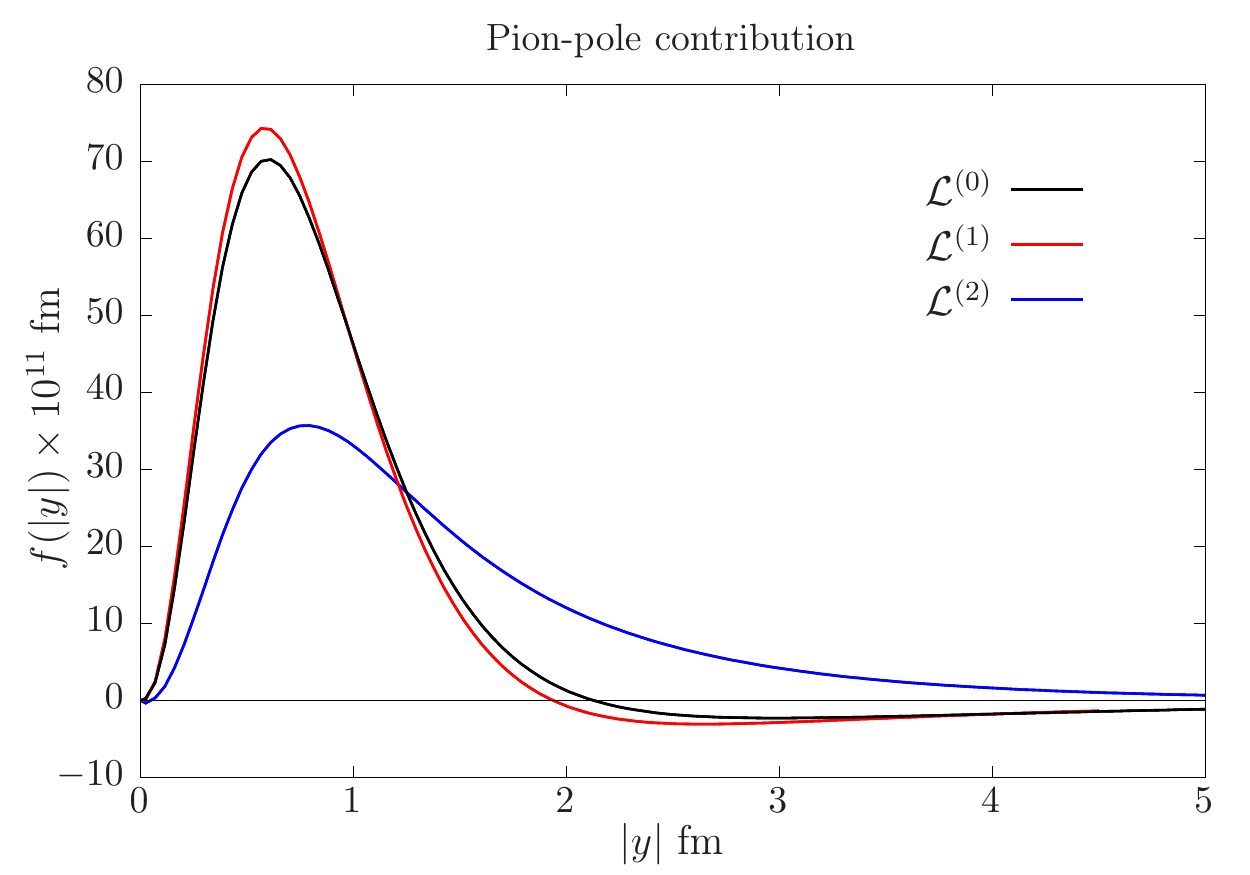}
\includegraphics[width=0.51\textwidth]{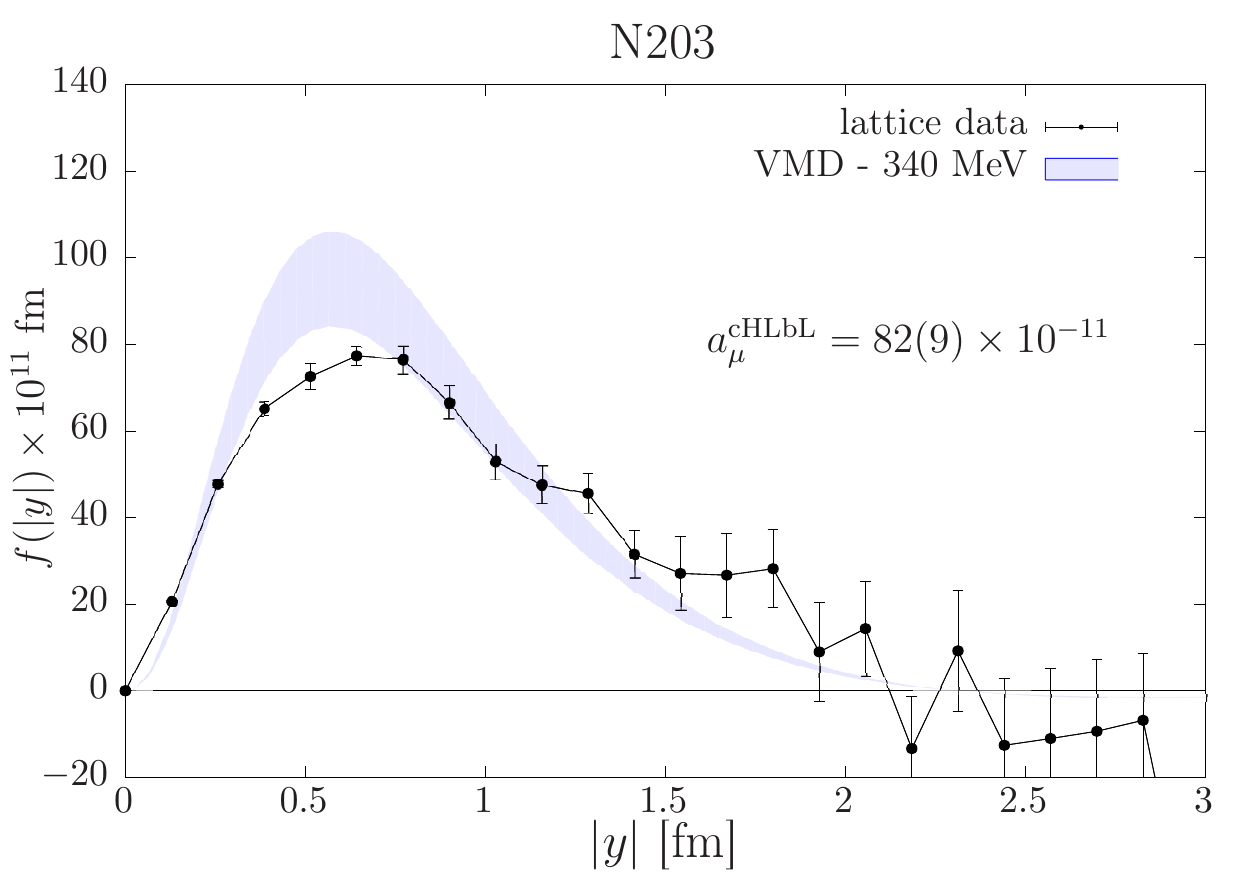}}
\caption{\la{fig:intgnd135CLS} \small
Left: 
Continuum, infinite-volume integrand of the final integral over $|y|$ to obtain the $\pi^0$ pole contribution to $\amuhlbl$ at
the physical mass $m_\pi=135\,$MeV,
assuming the VMD form of the transition form factor. The integrand differs, depending on the precise choice of 
QED kernel, without changing the result of the integral. The ${\cal L}^{(i)}$ correspond to subtracting different sets of 
$x$- or $y$-independent terms from the QED kernel.
Right: 
The integrand of the fully connected Wick-contraction diagrams 
yielding $\amuhlbl$ as a function of the integration variable $|y|$,
obtained on a $48^3\times96$ lattice at a pion mass of $m_\pi=340\,$MeV and lattice spacing $a=0.064\,$fm.
The kernel used is ${\cal L}^{(2)}$, which vanishes by construction whenever $x$ or $y$ vanishes.
The integrals over the vertices in $x$ and $z$ (see Fig.\ \ref{fig:HLbLkernel}) have already been performed at this stage. 
For comparison, the corresponding (continuum, infinite-volume) integrand for the $\pi^0$ pole contribution to $\amuhlbl$ is displayed, 
multiplied by an enhancement factor of 3 (lower edge of the band) and 34/9 (upper edge of the band) to account for the 
absence of the disconnected diagrams; see the end of section 2 for an explanation of these weight factors.
}
\end{figure}

In order to get an idea of the length scales involved in the problem,
one can inspect the coordinate-space dependence of the integrand for
the pion pole~\cite{Asmussen:2016lse}. An example is shown in
the left panel of Fig.\ \ref{fig:intgnd135CLS} for the physical pion mass. The curves
correspond to different kernels, which are equivalent in infinite
volume. They differ by $x$- or $y$-independent terms which do not
contribute to the integral, but modify the integrand. 
In~\cite{Blum:2017cer}, where these subtraction terms were first introduced,
it was shown that they can drastically reduce lattice discretization errors by forcing 
the kernel to vanish when some of the vertices coincide.
From the left panel of Fig.\ \ref{fig:intgnd135CLS}, it is clear that the integrand
is rather long-range in all three cases shown.

\subsection{Status of lattice results}

The calculation of hadronic light-by-light scattering in $(g-2)_\mu$
is still work in progress.  The most recent refereed publication by
the RBC/UKQCD collaboration~\cite{Blum:2016lnc} uses the QED$_L$
formulation and presents results for the Wick-contraction classes (4)
and (2+2) on a $48^3\times 96$ lattice at the physical pion mass with
a lattice spacing of $a^{-1}=1.73\,{\rm GeV}$.  The contribution of
the fully connected class of diagrams is $a_\mu^{{\rm HlbL}\,(4)} =
(116.0 \pm 9.6) \times 10^{-11}$, while the (2+2) diagrams yields
$a_\mu^{{\rm HlbL}\,(2+2)} =  (-62.5 \pm 8.0) \times 10^{-11}$.
Together, they amount to~\cite{Blum:2016lnc}
\be
\amuhlbl = (53.5\pm 13.5)\cdot 10^{-11},
\ee
where the quoted error is statistical.
This represents the first lattice result for the two leading
Wick-contraction topologies.  However, the authors acknowledge that
``The finite-volume and finite lattice-spacing errors could be quite
large and are the subject of ongoing research.'' We note that the
total is about a factor two lower than the model estimates.  A
possible explanation is that in the latter, neglected contributions
could be more important than so far expected; or, since the QED$_L$
method has O($1/L^2$) finite-size effects, the latter could be
responsible for a large systematic error. It was
noted~\cite{Gerardin:2017ryf} that, based on the model estimate and
large-$N_c$ inspired arguments, one would expect $a_\mu^{{\rm
    HlbL}\,(2+2)}$ to be dominated by the $(\pi^0,\eta,\eta')$ exchange
and approximately $-150\cdot 10^{-11}$ in size; the fully connected diagrams
would then have to amount to about $250\cdot 10^{-11}$ in order to recover 
the model result $\amuhlbl\approx 100\cdot 10^{-11}$ discussed in section \ref{sec:modeltest}.

\begin{figure}
\centerline{\includegraphics[width=0.33\textwidth]{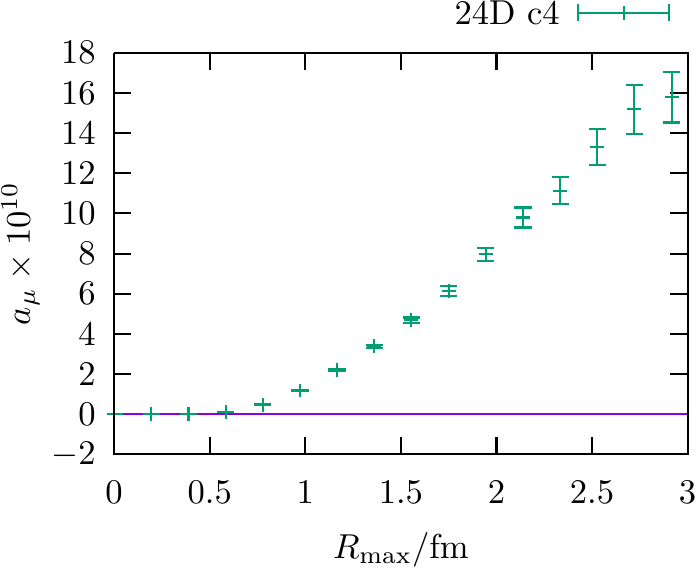}
\includegraphics[width=0.33\textwidth]{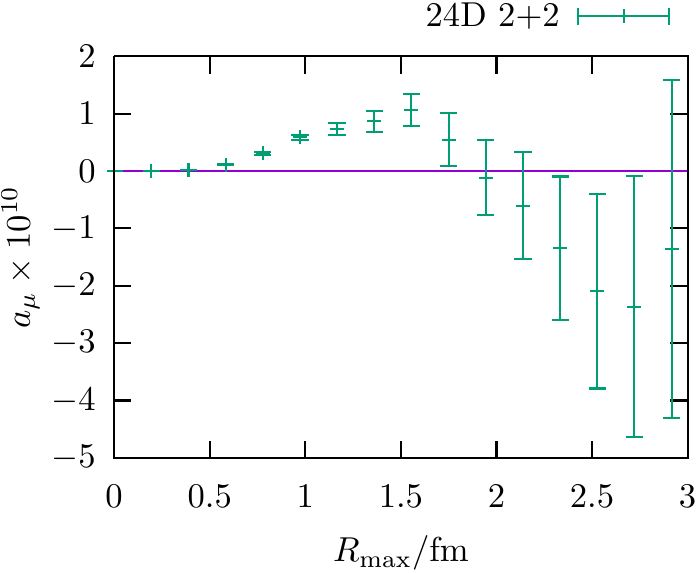}
\includegraphics[width=0.33\textwidth]{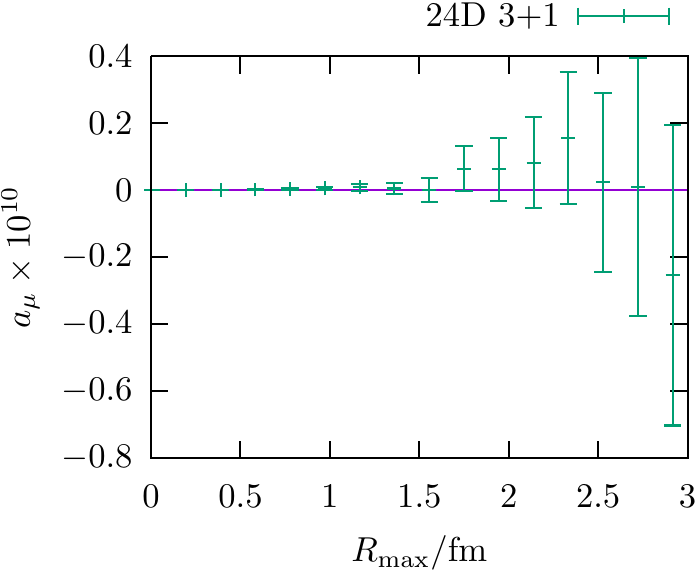}}
\caption{\la{fig:RBCprivcomm} Recent results by the RBC/UKQCD collaboration~\cite{JinPrivComm} obtained 
on a $24^3\times64$ lattice at the physical pion mass and $a^{-1}=1.015{\rm\, GeV}$.
The integral yielding  $\amuhlbl$ as a function of its upper limit, the integration variable being 
the maximum distance between any two internal vertices. The QED$_\infty$ formulation 
including subtractions~\cite{Blum:2017cer} was used. Separately (from left to right) 
for the Wick-contraction classes (4), (2+2) and (3+1). In the latter case, only those Wick-contractions are included 
in which the external photon is attached to the loop containing three vertices.}
\end{figure}

L.\ Jin presented an update of the RBC/UKQCD calculation at this year's 
lattice conference. An extrapolation to infinite volume and zero lattice spacing
based on several ensembles yielded the results $a_\mu^{{\rm HlbL}\,(4)}=(282\pm40)\cdot10^{-11}$
and $a_\mu^{{\rm HlbL}\,(2+2)}=(-163\pm34)\cdot 10^{-11}$, resulting in the sum 
$\amuhlbl = (119\pm53)\cdot10^{-11}$. These values are much more in line with the expectation
from the model calculations, but the extrapolation, and the cancellation between the two Wick-contraction
topologies, enhances the relative error on the final result.

Both the RBC/UKQCD collaboration and the Mainz group have started
generating lattice results with their respective QED$_\infty$ method.
As a very recent development, the RBC/UKQCD collaboration has performed a calculation of 
the three leading diagram topologies (4), (2+2) and (3+1) on a coarse lattice at the physical pion mass; see Fig.\ \ref{fig:RBCprivcomm}.
The (3+1) topology is found to make a negligible contribution~\cite{JinPrivComm}.
The Mainz group has computed and analyzed the fully connected set of diagrams (4)~\cite{Asmussen2018w}.
It uses rather fine lattices, with a typical lattice spacing of $a=0.064\,$fm, 
on the other hand the simulated pion masses ($m_\pi\gtrsim200\,$MeV) lie above the physical value. 
The integrand obtained at $m_\pi=340\,$MeV is displayed in the right panel of Fig.\ \ref{fig:intgnd135CLS},
and compared to the prediction corresponding to the neutral pion pole, including the appropriate enhancement weight factor,
as discussed at the end of section \ref{sec:modeltest}. The pion-pole integrand, predicted with a VMD transition form factor,
provides a surprisingly good approximation to the lattice data.
Tests of the systematic errors on the fully connected set of diagrams at $m_\pi=280\,$MeV are shown in
Fig.\ \ref{fig:fse_fle}: finite-size effects and discretization
effects appear to be under control. It remains to be seen how well controlled the extrapolation in the pion mass will be.
Restricting the sums over the vertex positions in a systematically controlled way to the regions that contribute appreciably is likely to
reduce the statistical noise.

\begin{figure}
\centerline{ \includegraphics[width=0.47\textwidth]{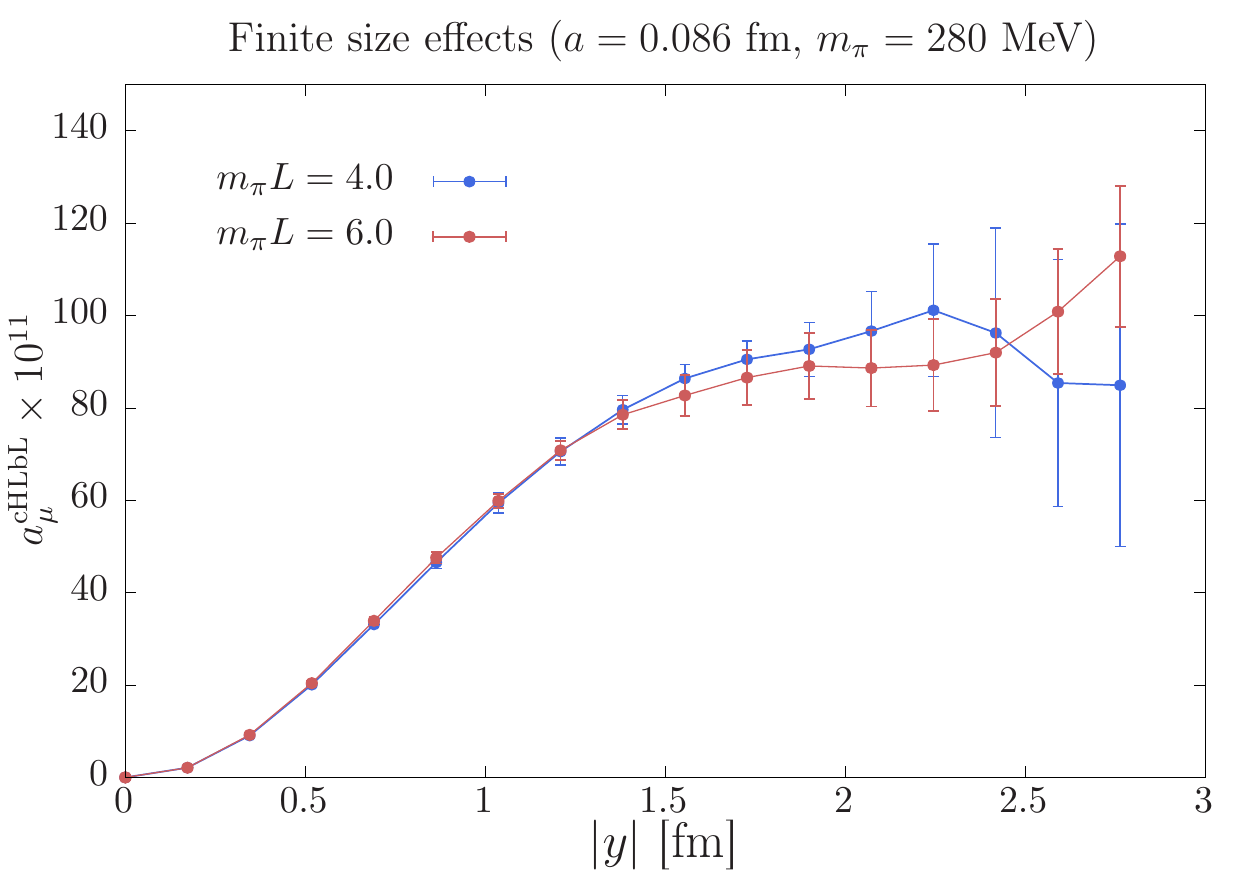}
 \includegraphics[width=0.47\textwidth]{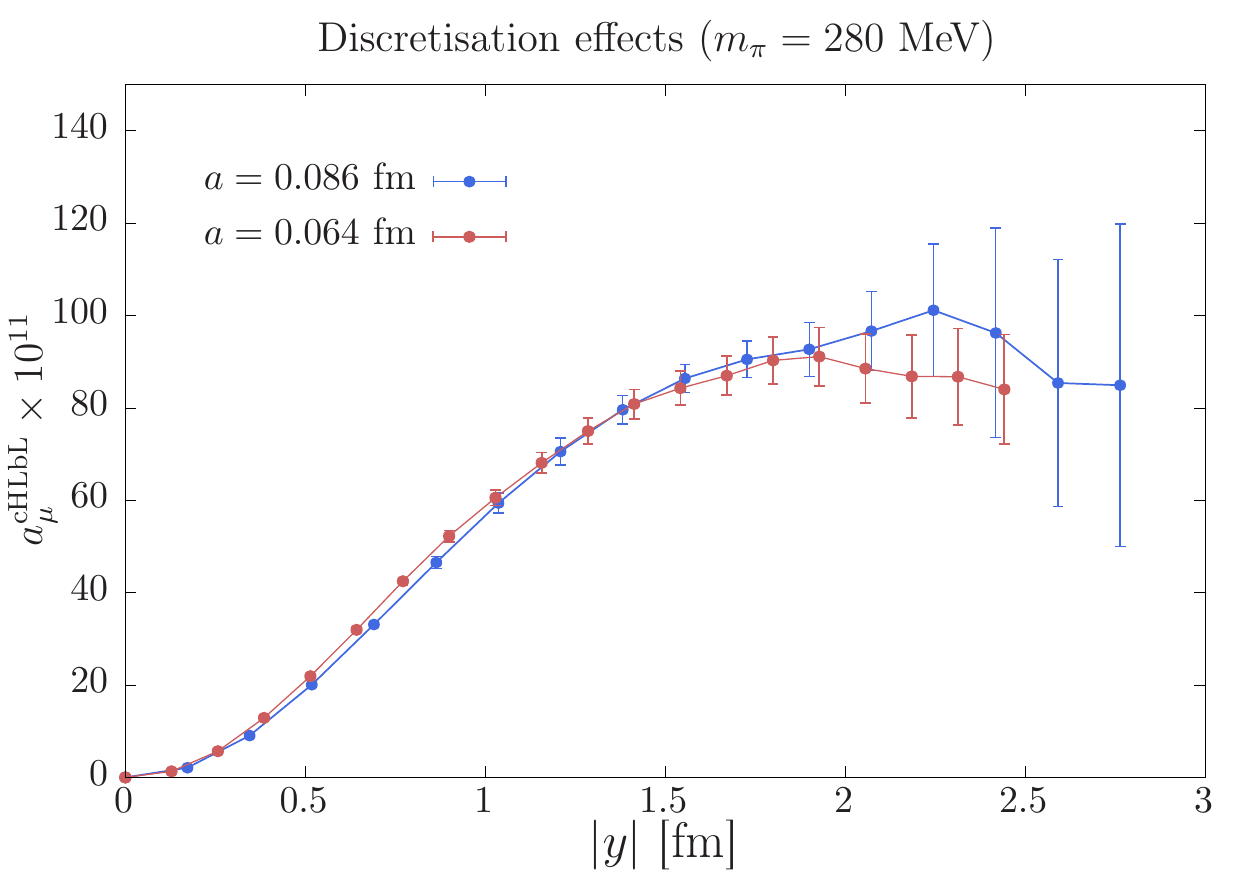}}
\caption{\la{fig:fse_fle} Mainz group: the integral yielding $\amuhlbl$ as a function of the upper limit of the integration variable $|y|$
at a pion mass of $m_\pi=280\,$MeV.
The integrals over the vertices in $x$ and $z$ have already been performed at this stage. 
The kernel used is ${\cal L}^{(2)}$, which vanishes by construction whenever $x$ or $y$ vanishes.
The left panel compares results from two different lattice volumes; the right panel compares two different lattice spacings.}
\end{figure}

\section{Conclusion}

The hadronic light-by-light scattering contribution is, along with the hadronic vacuum polarization contribution,
the leading source of uncertainty in the Standard Model prediction for the anomalous magnetic moment of the muon, $(g-2)_\mu$.
For decades, a framework which offers a systematically improvable prediction was lacking. This has now changed,
with significant progress having been made in the dispersive as well as lattice QCD approaches.
Even though model calculations will soon become superseded, valuable
lessons have been learnt from them, which can help control the
systematic errors in the ab initio approaches.

Given that the quantity $\amuhlbl$ involves three spacelike and one
quasi-real photon, lattice QCD is in a good position to provide a
first-principles prediction -- no analytic continuation is required.
At the same time, dealing with the four-point function of the vector
current is pushing the field into a territory on which the community
has little prior experience, hence dealing with the statistical and 
the systematic errors,
such as the finite lattice spacing and finite-volume errors, requires
special attention.  A cross-check between at least two independent
lattice collaborations is hence extremely valuable, as is a
comparison of the lattice results with the results of the dispersive
approaches.  In the latter case, it will be especially interesting to
see how the dispersive treatment of $\pi\pi$ intermediate states
differs quantitatively from previous estimates based on a
narrow (scalar and tensor) resonance exchange approximation.

\section*{Acknowledgements}
The research of the Mainz lattice QCD group is supported by the Deutsche Forschungsgemeinschaft (DFG) through 
the Collaborative Research Centre~1044 and the European Research Council (ERC)
under the European Union’s Horizon 2020 research and innovation
programme through grant agreement 771971-SIMDAMA.



\bibliography{/Users/harvey/BIBLIO/viscobib.bib}

\begin{thebibliography}{10}
\providecommand{\url}[1]{\texttt{#1}}
\providecommand{\urlprefix}{URL }
\expandafter\ifx\csname urlstyle\endcsname\relax
  \providecommand{\doi}[1]{doi:\discretionary{}{}{}#1}\else
  \providecommand{\doi}{doi:\discretionary{}{}{}\begingroup
  \urlstyle{rm}\Url}\fi
\providecommand{\eprint}[2][]{\url{#2}}

\bibitem{Bennett:2006fi}
G.~W. Bennett \emph{et~al.},
\newblock \emph{{Final Report of the Muon E821 Anomalous Magnetic Moment
  Measurement at BNL}},
\newblock Phys. Rev. \textbf{D73}, 072003 (2006),
\newblock \doi{10.1103/PhysRevD.73.072003},
\newblock \eprint{hep-ex/0602035}.

\bibitem{Jegerlehner:2009ry}
F.~Jegerlehner and A.~Nyffeler,
\newblock \emph{{The Muon g-2}},
\newblock Phys.Rept. \textbf{477}, 1 (2009),
\newblock \doi{10.1016/j.physrep.2009.04.003},
\newblock \eprint{0902.3360}.

\bibitem{Davier:2017zfy}
M.~Davier, A.~Hoecker, B.~Malaescu and Z.~Zhang,
\newblock \emph{{Reevaluation of the hadronic vacuum polarisation contributions
  to the Standard Model predictions of the muon $g-2$ and ${\alpha (m_Z^2)}$
  using newest hadronic cross-section data}},
\newblock Eur. Phys. J. \textbf{C77}(12), 827 (2017),
\newblock \doi{10.1140/epjc/s10052-017-5161-6},
\newblock \eprint{1706.09436}.

\bibitem{Fertl:2016nij}
M.~Fertl,
\newblock \emph{{Next generation muon g-2 experiment at FNAL}},
\newblock Hyperfine Interact. \textbf{237}(1), 94 (2016),
\newblock \doi{10.1007/s10751-016-1304-7},
\newblock \eprint{1610.07017}.

\bibitem{Otani:2015lra}
M.~Otani,
\newblock \emph{{Design of the J-PARC MUSE H-line for the Muon g-2/EDM
  Experiment at J-PARC (E34)}},
\newblock JPS Conf. Proc. \textbf{8}, 025010 (2015),
\newblock \doi{10.7566/JPSCP.8.025010}.

\bibitem{Keshavarzi:2018mgv}
A.~Keshavarzi, D.~Nomura and T.~Teubner,
\newblock \emph{{Muon $g-2$ and $\alpha(M_Z^2)$: a new data-based analysis}},
\newblock Phys. Rev. \textbf{D97}(11), 114025 (2018),
\newblock \doi{10.1103/PhysRevD.97.114025},
\newblock \eprint{1802.02995}.

\bibitem{Meyer:2018til}
H.~B. Meyer and H.~Wittig,
\newblock \emph{{Lattice QCD and the anomalous magnetic moment of the muon}}
  (2018),
\newblock \eprint{1807.09370}.

\bibitem{Jegerlehner:2018gjd}
F.~Jegerlehner,
\newblock \emph{{The Role of Mesons in Muon $g-2$}}  (2018),
\newblock \eprint{1809.07413}.

\bibitem{Prades:2009tw}
J.~Prades, E.~de~Rafael and A.~Vainshtein,
\newblock \emph{{The Hadronic Light-by-Light Scattering Contribution to the
  Muon and Electron Anomalous Magnetic Moments}},
\newblock Adv. Ser. Direct. High Energy Phys. \textbf{20}, 303 (2009),
\newblock \doi{10.1142/9789814271844_0009},
\newblock \eprint{0901.0306}.

\bibitem{Knecht:2001qf}
M.~Knecht and A.~Nyffeler,
\newblock \emph{{Hadronic light by light corrections to the muon g-2: The Pion
  pole contribution}},
\newblock Phys.Rev. \textbf{D65}, 073034 (2002),
\newblock \doi{10.1103/PhysRevD.65.073034},
\newblock \eprint{hep-ph/0111058}.

\bibitem{Knecht:2001qg}
M.~Knecht, A.~Nyffeler, M.~Perrottet and E.~de~Rafael,
\newblock \emph{{Hadronic light by light scattering contribution to the muon
  g-2: An Effective field theory approach}},
\newblock Phys. Rev. Lett. \textbf{88}, 071802 (2002),
\newblock \doi{10.1103/PhysRevLett.88.071802},
\newblock \eprint{hep-ph/0111059}.

\bibitem{Landau:1948kw}
L.~D. Landau,
\newblock \emph{{On the angular momentum of a system of two photons}},
\newblock Dokl. Akad. Nauk Ser. Fiz. \textbf{60}(2), 207 (1948),
\newblock \doi{10.1016/B978-0-08-010586-4.50070-5}.

\bibitem{Yang:1950rg}
C.-N. Yang,
\newblock \emph{{Selection Rules for the Dematerialization of a Particle Into
  Two Photons}},
\newblock Phys. Rev. \textbf{77}, 242 (1950),
\newblock \doi{10.1103/PhysRev.77.242}.

\bibitem{Green:2015sra}
J.~Green, O.~Gryniuk, G.~von Hippel, H.~B. Meyer and V.~Pascalutsa,
\newblock \emph{{Lattice QCD calculation of hadronic light-by-light
  scattering}},
\newblock Phys. Rev. Lett. \textbf{115}(22), 222003 (2015),
\newblock \doi{10.1103/PhysRevLett.115.222003},
\newblock \eprint{1507.01577}.

\bibitem{Gerardin:2017ryf}
A.~G{\'e}rardin, J.~Green, O.~Gryniuk, G.~von Hippel, H.~B. Meyer,
  V.~Pascalutsa and H.~Wittig,
\newblock \emph{{Hadronic light-by-light scattering amplitudes from lattice QCD
  versus dispersive sum rules}},
\newblock Phys. Rev. \textbf{D98}(7), 074501 (2018),
\newblock \doi{10.1103/PhysRevD.98.074501},
\newblock \eprint{1712.00421}.

\bibitem{Pascalutsa:2010sj}
V.~Pascalutsa and M.~Vanderhaeghen,
\newblock \emph{{Sum rules for light-by-light scattering}},
\newblock Phys.Rev.Lett. \textbf{105}, 201603 (2010),
\newblock \doi{10.1103/PhysRevLett.105.201603},
\newblock \eprint{1008.1088}.

\bibitem{Pascalutsa:2012pr}
V.~Pascalutsa, V.~Pauk and M.~Vanderhaeghen,
\newblock \emph{{Light-by-light scattering sum rules constraining meson
  transition form factors}},
\newblock Phys.Rev. \textbf{D85}, 116001 (2012),
\newblock \doi{10.1103/PhysRevD.85.116001},
\newblock \eprint{1204.0740}.

\bibitem{Gerardin:2016cqj}
A.~Gerardin, H.~B. Meyer and A.~Nyffeler,
\newblock \emph{{Lattice calculation of the pion transition form factor $\pi^0
  \to \gamma^* \gamma^*$}},
\newblock Phys. Rev. \textbf{D94}(7), 074507 (2016),
\newblock \doi{10.1103/PhysRevD.94.074507},
\newblock \eprint{1607.08174}.

\bibitem{Bijnens:2016hgx}
J.~Bijnens and J.~Relefors,
\newblock \emph{{Pion light-by-light contributions to the muon $g-2$}},
\newblock JHEP \textbf{09}, 113 (2016),
\newblock \doi{10.1007/JHEP09(2016)113},
\newblock \eprint{1608.01454}.

\bibitem{Pauk:2014rfa}
V.~Pauk and M.~Vanderhaeghen,
\newblock \emph{{Anomalous magnetic moment of the muon in a dispersive
  approach}},
\newblock Phys.Rev. \textbf{D90}(11), 113012 (2014),
\newblock \doi{10.1103/PhysRevD.90.113012},
\newblock \eprint{1409.0819}.

\bibitem{Colangelo:2014dfa}
G.~Colangelo, M.~Hoferichter, M.~Procura and P.~Stoffer,
\newblock \emph{{Dispersive approach to hadronic light-by-light scattering}},
\newblock JHEP \textbf{09}, 091 (2014),
\newblock \doi{10.1007/JHEP09(2014)091},
\newblock \eprint{1402.7081}.

\bibitem{Hagelstein:2017obr}
F.~Hagelstein and V.~Pascalutsa,
\newblock \emph{{Dissecting the Hadronic Contributions to $(g-2)_\mu$ by
  Schwinger’s Sum Rule}},
\newblock Phys. Rev. Lett. \textbf{120}(7), 072002 (2018),
\newblock \doi{10.1103/PhysRevLett.120.072002},
\newblock \eprint{1710.04571}.

\bibitem{Colangelo:2015ama}
G.~Colangelo, M.~Hoferichter, M.~Procura and P.~Stoffer,
\newblock \emph{{Dispersion relation for hadronic light-by-light scattering:
  theoretical foundations}},
\newblock JHEP \textbf{09}, 074 (2015),
\newblock \doi{10.1007/JHEP09(2015)074},
\newblock \eprint{1506.01386}.

\bibitem{Colangelo:2017qdm}
G.~Colangelo, M.~Hoferichter, M.~Procura and P.~Stoffer,
\newblock \emph{{Rescattering effects in the hadronic-light-by-light
  contribution to the anomalous magnetic moment of the muon}},
\newblock Phys. Rev. Lett. \textbf{118}(23), 232001 (2017),
\newblock \doi{10.1103/PhysRevLett.118.232001},
\newblock \eprint{1701.06554}.

\bibitem{Colangelo:2017fiz}
G.~Colangelo, M.~Hoferichter, M.~Procura and P.~Stoffer,
\newblock \emph{{Dispersion relation for hadronic light-by-light scattering:
  two-pion contributions}},
\newblock JHEP \textbf{04}, 161 (2017),
\newblock \doi{10.1007/JHEP04(2017)161},
\newblock \eprint{1702.07347}.

\bibitem{Colangelo2018w}
G.~Colangelo,
\newblock \emph{{Talk at Second Plenary Workshop of the Muon g-2 Theory
  Initiative}}  (Mainz, 18-22 June 2018).

\bibitem{Danilkin:2018qfn}
I.~Danilkin and M.~Vanderhaeghen,
\newblock \emph{{Dispersive analysis of the $\gamma\gamma^{*} \to \pi \pi$
  process}}  (2018),
\newblock \eprint{1810.03669}.

\bibitem{Hoferichter:2018dmo}
M.~Hoferichter, B.-L. Hoid, B.~Kubis, S.~Leupold and S.~P. Schneider,
\newblock \emph{{Pion-pole contribution to hadronic light-by-light scattering
  in the anomalous magnetic moment of the muon}},
\newblock Phys. Rev. Lett. \textbf{121}(11), 112002 (2018),
\newblock \doi{10.1103/PhysRevLett.121.112002},
\newblock \eprint{1805.01471}.

\bibitem{Gerardin2018w}
A.~G\'erardin,
\newblock \emph{{Talk at Second Plenary Workshop of the Muon g-2 Theory
  Initiative}}  (Mainz, 18-22 June 2018).

\bibitem{Hayakawa:2005eq}
M.~Hayakawa, T.~Blum, T.~Izubuchi and N.~Yamada,
\newblock \emph{{Hadronic light-by-light scattering contribution to the muon
  g-2 from lattice QCD: Methodology}},
\newblock PoS \textbf{LAT2005}, 353 (2006),
\newblock \eprint{hep-lat/0509016}.

\bibitem{Blum:2015gfa}
T.~Blum, N.~Christ, M.~Hayakawa, T.~Izubuchi, L.~Jin and C.~Lehner,
\newblock \emph{{Lattice Calculation of Hadronic Light-by-Light Contribution to
  the Muon Anomalous Magnetic Moment}},
\newblock Phys. Rev. \textbf{D93}(1), 014503 (2016),
\newblock \doi{10.1103/PhysRevD.93.014503},
\newblock \eprint{1510.07100}.

\bibitem{Blum:2017cer}
T.~Blum, N.~Christ, M.~Hayakawa, T.~Izubuchi, L.~Jin, C.~Jung and C.~Lehner,
\newblock \emph{{Using infinite volume, continuum QED and lattice QCD for the
  hadronic light-by-light contribution to the muon anomalous magnetic moment}},
\newblock Phys. Rev. \textbf{D96}(3), 034515 (2017),
\newblock \doi{10.1103/PhysRevD.96.034515},
\newblock \eprint{1705.01067}.

\bibitem{Green:2015mva}
J.~Green, N.~Asmussen, O.~Gryniuk, G.~von Hippel, H.~B. Meyer, A.~Nyffeler and
  V.~Pascalutsa,
\newblock \emph{{Direct calculation of hadronic light-by-light scattering}},
\newblock PoS \textbf{LATTICE2015}, 109 (2016),
\newblock \eprint{1510.08384}.

\bibitem{Asmussen:2016lse}
N.~Asmussen, J.~Green, H.~B. Meyer and A.~Nyffeler,
\newblock \emph{{Position-space approach to hadronic light-by-light scattering
  in the muon $g-2$ on the lattice}},
\newblock PoS \textbf{LATTICE2016}, 164 (2016),
\newblock \doi{10.22323/1.256.0164},
\newblock \eprint{1609.08454}.

\bibitem{Blum:2016lnc}
T.~Blum, N.~Christ, M.~Hayakawa, T.~Izubuchi, L.~Jin, C.~Jung and C.~Lehner,
\newblock \emph{{Connected and Leading Disconnected Hadronic Light-by-Light
  Contribution to the Muon Anomalous Magnetic Moment with a Physical Pion
  Mass}},
\newblock Phys. Rev. Lett. \textbf{118}(2), 022005 (2017),
\newblock \doi{10.1103/PhysRevLett.118.022005},
\newblock \eprint{1610.04603}.

\bibitem{JinPrivComm}
L.~Jin,
\newblock \emph{{Private communication}}  (2018).

\bibitem{Asmussen2018w}
N.~Asmussen,
\newblock \emph{{Talk at Second Plenary Workshop of the Muon g-2 Theory
  Initiative}}  (Mainz, 18-22 June 2018).

\end{thebibliography}

\nolinenumbers

\end{document}